\documentclass[aps,prb,showpacs,twocolumn]{revtex4-1}
\usepackage{amssymb}
\usepackage{amsmath}
\usepackage{graphicx}
\usepackage{epsfig}

\begin{document}

\title{Anderson localization induced by complex potential}
\author{R. Wang, K. L. Zhang, and Z. Song}
\email{songtc@nankai.edu.cn}
\affiliation{School of Physics, Nankai University, Tianjin 300071, China}

\begin{abstract}
Uncorrelated disorder potential in one-dimensional lattice definitely
induces Anderson localization, while quasiperiodic potential can lead to
both localized and extended phases, depending on the potential strength. We
investigate the Anderson localization in one-dimensional lattice\ with
non-Hermitian complex disorder and quasiperiodic potential. We present a
non-Hermitian \textrm{SSH} chain to demonstrate that a nonzero imaginary
disorder on-site potential can induce the standard Anderson localization. We
also show that the non-Hermitian Aubry-Andre (\textrm{AA}) model exhibits a
transition in parameter space, which separates the localization and
delocalization phases and is determined by the self-duality of the model. It
indicates that a pure imaginary quasiperiodic potential takes the same role
as a real quasiperiodic potential in the transition point between
localization and delocalization. Remarkably, a system with complex
quasiperiodic potential exhibits interference-like pattern on the transition
points.
\end{abstract}

\maketitle

\section{Introduction}

The localization phase in quantum systems, which is originally rooted in
condensed matter \cite{PA}, has recently attracted a lot of theoretical and
experimental interest in a variety of fields, including light waves in
optical random media \cite{ST,SG,CC,DM}, matter waves in optical potential
\cite{KD,GR,FJ,GS}, sound waves in elastic media \cite{HH}, and quantum
chaotic systems\cite{JC}, since the localization of quantum particles could
prevent the transport necessary for equilibration in isolated systems.
Anderson localization\cite{PW} predicts that single-particle wave functions
become localized in the presence of some uncorrelated disorder, leading to a
metal-insulator transition caused by the quantum interference in the
scattering processes of a particle with random impurities and defects. A
conventional Anderson localization is not controllable in one and
two-dimensional systems. Nevertheless, localization does not require
disorder and fortunately, the Aubry-Andre model \cite{MY,SA}, which has
quasiperiodic potential, exhibits a transition between a localized and
extended phases. In practice, quasiperiodic potential arises naturally in
optical experiments using lasers with incommensurate wave vectors.
Accordingly, many experiments in such systems have now observed
single-particle localization \cite{LDN,LF,GR,YL,GM,MS}. While any optical
system that includes gain or loss is non-Hermitian by nature, it has been
shown that the presence of the imaginary potential causes many surprising
effects \cite{JLPRL,AAClerk,Koutserimpas,HRamezani,Huang}.

Motivated by the recent development of non-Hermitian quantum mechanics \cite%
{BenderRPP} both in theoretical and experimental aspects \cite%
{PRL08a,PRL08b,Klaiman,CERuter,YDChong,Regensburger,LFeng,Fleury,NM,FL,Ganainy18,YFChen,Christodoulides}%
, we investigate the localization transitions in non-Hermitian regime.
Several pioneer works have been devoted to investigate the tight-binding
system with non-Hermition $\mathcal{PT}$-symmetric quasiperiodic potential
\cite{CY,SL1,SL2}. The extension from real potential to a complex one raises
the question of whether the real and imaginary part of complex potential has
correlated effect on the Anderson localization. Especially, since the
quasiperiodic potential does possess an intrinsic phase, the phase
difference of real and imaginary quasiperiodic potential may influence the
Anderson localization transition. This is the main purpose of the present
work. We investigate the Anderson localization in one-dimensional lattice
with non-Hermitian complex disorder and quasiperiodic potential. We present
a non-Hermitian \textrm{SSH} chain to demonstrate that a nonzero imaginary
disorder on-site potential can induce the standard Anderson localization. We
also show that the non-Hermitian Aubry-Andre (\textrm{AA}) model exhibits a
transition in parameter space, which separates the localization and
delocalization phases and is determined by the self-duality of the model. It
indicates that a pure imaginary quasiperiodic potential takes the same role
as a real quasiperiodic potential in the transition point between
localization and delocalization. Remarkably, a system with complex
quasiperiodic potential exhibits interference-like pattern on the transition
point, i.e., the phase difference between real and imaginary quasiperiodic
potential determines the boundary of transition. Our approach opens a new
way to investigate the interplay of localization and gain/loss in
non-Hermitian system.

This paper is organized as follows. In Sec. \ref{Localization in
non-Hermitian SSH chain}, we present a non-Hermitian \textrm{SSH} chain with
disorder staggered balanced gain and loss. We map this model to an
equivalent Hermitian one and show the existence of Anderson localization.\
In Sec. \ref{Self-duality under imaginary potential}, we show exactly that a
non-Hermitian \textrm{AA} Hamiltonian with imaginary quasi-periodic
potential possesses a self-duality and manifests a localization transition
at the self-dual point. In Sec. \ref{Interference effect on localization
transition}, we demonstrate that the real and imaginary quasiperiodic
potential has interference effect on the Anderson localization transition.
Finally, we give a summary in Section \ref{Conclusion}.

\section{Localization in non-Hermitian SSH chain}

\label{Localization in non-Hermitian SSH chain}

\begin{figure}[tbph]
\includegraphics[ bb=30 365 478 761, width=0.4\textwidth, clip]{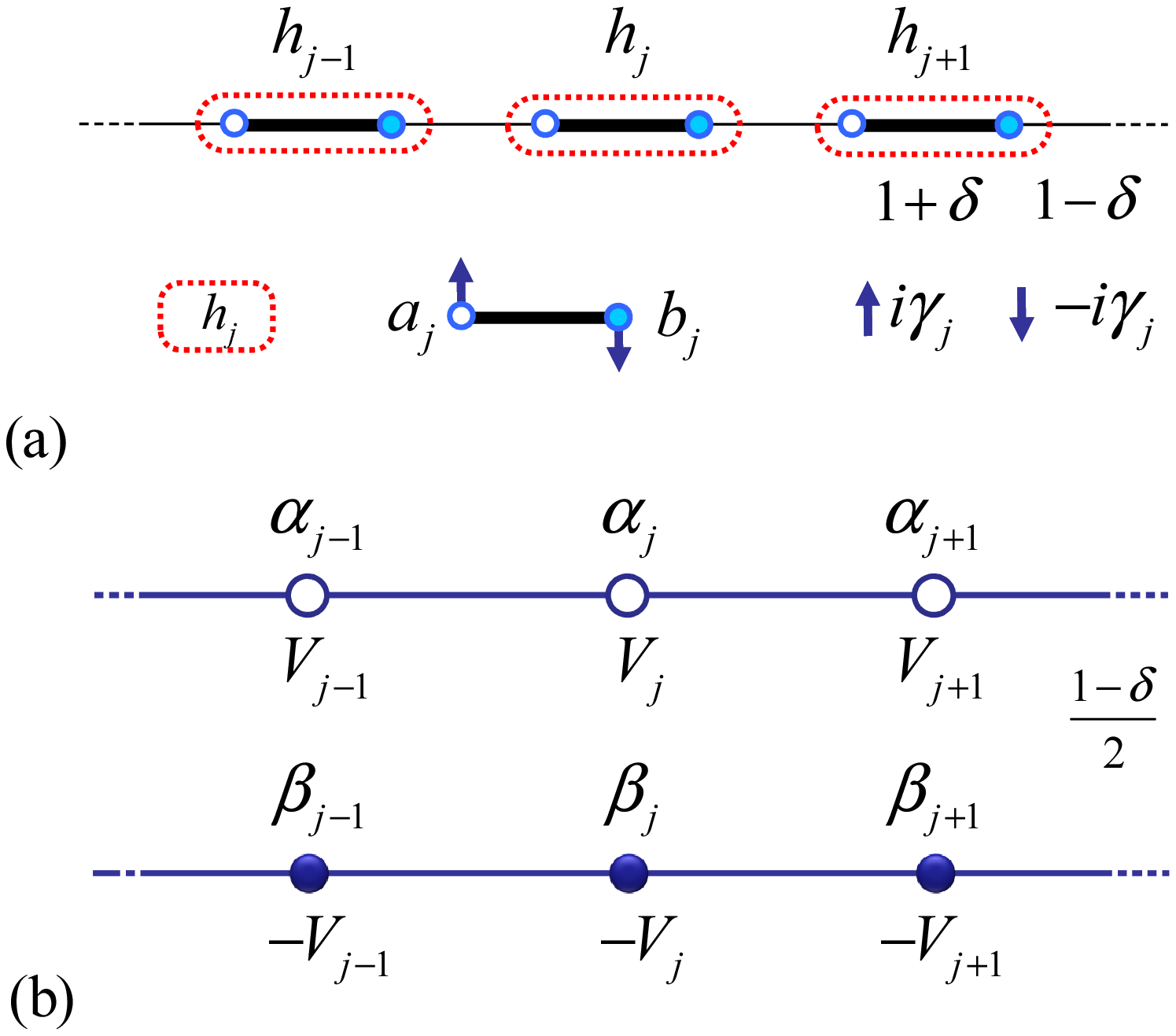}
\caption{(Color online) Schematic illustration of the equivalence between
imaginary and real potential via non-Hermitian \textrm{SSH} chain in strong
dimerization limit. (a) Lattice geometry for the model described in Eq. (%
\protect\ref{SSH}), that represents a non-Hermitian \textrm{SSH} chain with
position-dependent $\mathcal{PT}$-symmetric imaginary potential pairs. Solid
(empty) circle indicates A (B) lattice, while thin and thick solid lines
indicate distorted hopping terms. A $\mathcal{PT}$ dimmer (circled by the
red dashed line) contains two strongly coupled sites with opposite imaginary
potential (indicated by opposite arrows). (b) Equivalent model described in
Eq. (\protect\ref{SSH2}), that represents two independent uniform chains
with opposite position-dependent real potential. If the imaginary potential $%
i\protect\gamma _{j}$ in (a) is disorder, the real potential $V_{j}$ in (b)
is also random distribution, supporting Anderson localization.}
\label{fig1}
\end{figure}

We consider a non-Hermitian \textrm{SSH} chain with disorder staggered
balanced gain and loss. The simplest tight-binding model with these features
is
\begin{eqnarray}
H_{\mathrm{SSH}} &=&(1+\delta )\sum_{j=1}^{N}a_{j}^{\dag }b_{j}+(1-\delta
)\sum_{j=1}^{N-1}b_{j}^{\dag }a_{j+1}  \notag \\
&&+\mathrm{H.c.}+i\sum_{j=1}^{N}\gamma _{j}(a_{j}^{\dag }a_{j}-b_{j}^{\dag
}b_{j}),  \label{SSH}
\end{eqnarray}%
where $\delta $ and $\gamma _{j}$, are the distortion factor with unit
tunneling constant and the alternating imaginary potential magnitude at
dimmer $j$, respectively. In this work, we focus on the weak limit of
imaginary potential $\left\vert \gamma _{j}\right\vert \ll 1$. Here $%
a_{l}^{\dag }$ and $b_{l}^{\dag }$ are the creation operator of the particle
at the $l$th site in $A$ and $B$\ sub-lattices. The particle can be fermion
or boson, depending on their own commutation relations. (A sketch of the
lattice has been shown in Fig. \ref{fig1}.) In the following\textbf{, }we
will show that a nonzero disorder staggered balanced gain and loss can lead
to standard Anderson localization.

We start with the non-Hermitian $\mathcal{PT}$-symmetric dimmer at position $%
j$, with the Hamiltonian
\begin{equation}
h_{j}=(1+\delta )(a_{j}^{\dag }b_{j}+\mathrm{H.c.})+i\gamma _{j}(a_{j}^{\dag
}a_{j}-b_{j}^{\dag }b_{j}).
\end{equation}%
It can be diagonalized as the form
\begin{equation}
h_{j}=\sqrt{(1+\delta )^{2}-\gamma _{j}^{2}}(\overline{\alpha }_{j}\alpha
_{j}-\overline{\beta }_{j}\beta _{j}),
\end{equation}%
by introducing particle operators \cite{JL1,JL2}
\begin{equation}
\left\{
\begin{array}{cc}
\alpha _{j}=\frac{a_{j}+e^{-i\varphi _{j}}b_{j}}{1+ie^{-i\varphi _{j}}}, &
\beta _{j}=\frac{a_{j}-e^{i\varphi _{j}}b_{j}}{1+ie^{i\varphi _{j}}} \\
\overline{\alpha }_{j}=\frac{a_{j}^{\dag }+e^{-i\varphi _{j}}b_{j}^{\dag }}{%
1-ie^{-i\varphi _{j}}}, & \overline{\beta }_{j}=\frac{a_{j}^{\dag
}-e^{i\varphi _{j}}b_{j}^{\dag }}{1-ie^{i\varphi _{j}}}%
\end{array}%
\right. ,
\end{equation}%
with
\begin{equation}
\tan \varphi _{j}=\frac{\gamma _{j}}{\sqrt{(1+\delta )^{2}-\gamma _{j}^{2}}}.
\end{equation}%
Based on the identity
\begin{eqnarray}
&&(1+\delta )(a_{j}^{\dag }b_{j}+b_{j}^{\dag }a_{j})+i\gamma
_{j}(a_{j}^{\dag }a_{j}-b_{j}^{\dag }b_{j})  \notag \\
&=&\sqrt{(1+\delta )^{2}-\gamma _{j}^{2}}(\overline{\alpha }_{j}\alpha _{j}-%
\overline{\beta }_{j}\beta _{j}),
\end{eqnarray}%
and the condition
\begin{equation}
1+\delta \gg 1-\delta ,\left\vert \gamma _{j}\right\vert \ll 1,
\end{equation}%
which lead to $\cos \varphi _{j}\approx e^{i\varphi _{j}}\approx 1$, one can
neglect the transition terms between sites with opposite potential, and get
the approximate expression
\begin{eqnarray}
H_{\mathrm{SSH}} &\approx &\frac{1}{2}(1-\delta )\sum_{j=1}^{N-1}\left(
\overline{\alpha }_{j}\alpha _{j+1}-\overline{\beta }_{j}\beta _{j+1}\right)
\notag \\
&&+\mathrm{H.c.}+\sum_{j=1}^{N}V_{j}(\overline{\alpha }_{j}\alpha _{j}-%
\overline{\beta }_{j}\beta _{j}).  \label{SSH2}
\end{eqnarray}%
The original system reduces to two independent uniform chains with opposite
real on-site random potential $V_{j}=\sqrt{(1+\delta )^{2}-\gamma _{j}^{2}}$%
. Hamiltonian $H_{\mathrm{SSH}}$ is diagonalizable since operators $\left(
\alpha _{j},\beta _{j},\overline{\alpha }_{j},\overline{\beta }_{j}\right) $
satisfy the canonical commutation relations. It indicates that $H_{\mathrm{%
SSH}}$\ can be regarded as a Hermitian model in the context of biorthogonal
inner product. On the other hand, we have shown that if a state
\begin{equation}
\left\vert \psi \right\rangle =\sum_{j}(A_{j}\alpha _{j}^{\dag }\left\vert
0\right\rangle +B_{j}\beta _{j}^{\dag }\left\vert 0\right\rangle ),
\end{equation}%
\ with $A_{j}B_{j}=0$ for all $j$, i.e., it has only single-chain component,
$\left\{ \alpha _{j}^{\dag }\left\vert 0\right\rangle \right\} $ or $\left\{
\beta _{j}^{\dag }\left\vert 0\right\rangle \right\} $, the dynamics of $%
\left\vert \psi \right\rangle $\ is the same as that in a Hermitian chain,
exhibiting the standard Anderson localization. Furthermore, the mapping
between the Hermitian and non-Hermitian Hamiltonian matrices (i.e., the
similar matrix) is a local transformation, which cannot result in
localization transition. In conclusion, we provide an example to demonstrate
that a nonzero imaginary disorder on-site potential can induce the Anderson
localization.

Numerical simulation is performed to demonstrate our conclusion. For
simplicity, we consider a uniform chain with disorder potential. The
Hamiltonian takes the form
\begin{equation}
H_{\text{\textrm{And}}}=\sum_{l}a_{l}^{\dag }a_{l+1}+\mathrm{H.c.}%
+\sum_{l}\mu _{l}a_{l}^{\dag }a_{l},
\end{equation}%
where the potential is simulated by a sequence of random complex variables.\
According to the theory of Anderson localization, all the eigenstates are
localized for any nonzero real random variables. On the other hand, our
analysis indicates that imaginary random variables may also induce localized
states.\ We perform the simulation for two cases, (i) $\mu _{l}=$ \textrm{ran%
}($-b,b$)$e^{i\phi },$ with $\phi =0$, $\pi /4$, and $\pi /2$, (ii) $\mu
_{l}=\mu _{0}e^{i\text{\textrm{ran(}}-\pi ,\pi \text{\textrm{)}}}$. Here $%
\mathrm{ran}(-b,b)$\ denotes a uniform random number within $(-b,b)$. In
case (i) with $\phi =0$, it corresponds to real disorder potential, which is
in the framework of Anderson localization. In case (i) with $\phi =\pi /2$,
it corresponds to imaginary disorder potential, which is the situation of
our prediction. In case (i) with $\phi =\pi /4$, the real and imaginary
disorder potential has the same amplitude. In case (ii), the real and
imaginary part of potential is out of phase but keep an identical norm.

\begin{figure*}[tbph]
\includegraphics[width=1\textwidth, clip]{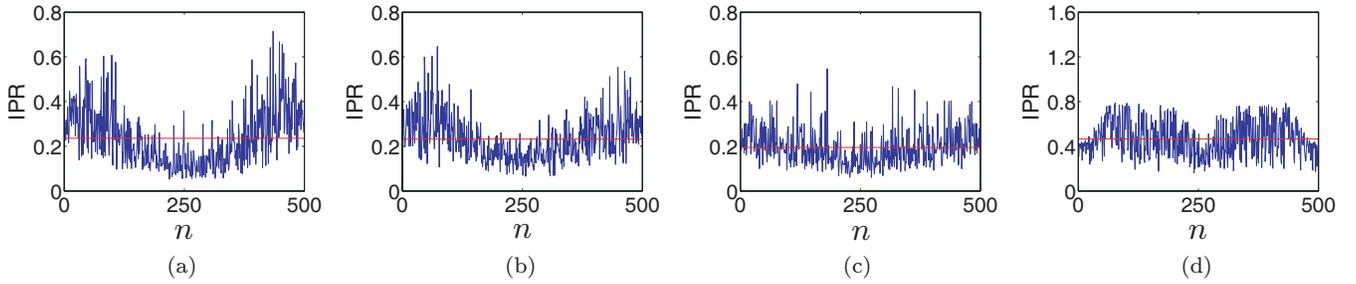}
\caption{(Color online) Numerical simulations for a uniform ring model with
random potential sequence $\left\{ u_{l}\right\} $. Plots of \textrm{IPR}
(blue line) along the energy level index $n$ for four typical sequence $%
\left\{ u_{l}\right\} $, which have the expression $\protect\mu _{l}=$%
\textrm{ran(}$-b,b$\textrm{)}$e^{i\protect\phi }$ with (a) $\protect\phi =0$%
, (b) $\protect\phi =\frac{\protect\pi }{4}$, (c) $\protect\phi =\frac{%
\protect\pi }{2}$, and (d) $\protect\mu _{l}=\protect\mu _{0}e^{i\text{%
\textrm{ran(}}-\protect\pi ,\protect\pi \text{\textrm{)}}}$. The value of
\textrm{AIPR} is described by the red line. The system parameters are $N=500$%
, $u_{0}=2$, $b=2$. It shows that (a), (b), and (c) have similar results,
indicating that a complex disorder potential takes the same role as a real
one to form Anderson localization. Panel (d) shows a different pattern,
which may be due to the different type of expression. Nevertheless the
result still accords with our prediction.}
\label{fig2}
\end{figure*}

To characterize the localization natures, we employ the inverse
participation ratio (\textrm{IPR}), as a criterion to distinguish the
extended states from the localized ones, which is defined as
\begin{equation}
\mathrm{IPR}^{(n)}=\frac{\sum_{l}\left\vert \left\langle \psi
_{n}\right\vert l\rangle \right\vert ^{4}}{\left( \sum_{l}\left\vert
\left\langle \psi _{n}\right\vert l\rangle \right\vert ^{2}\right) ^{2}}.
\end{equation}%
Meanwhile, the average \textrm{IPR} (\textrm{AIPR}) for all energy levels is
defined as the form of \textrm{AIPR}$=\frac{1}{N}\sum_{n=1}^{N}$\textrm{IPR}$%
^{(n)}$.\ For spatially extended states, it approaches to zero, whereas it
is finite for localized states. For a non-Hermitian system we introduce the
quantity, dressed energy
\begin{equation}
\tilde{\varepsilon}_{n}=\mathrm{sgn}[\text{\textrm{real}}(\varepsilon
_{n})]|\varepsilon _{n}|
\end{equation}%
to specifies the energy in non-Hermitian regime. We plot the result in Fig. %
\ref{fig2}, which show that the complex random potential of several types
can induce Anderson localization as our prediction.

\section{Self-duality under imaginary potential}

\label{Self-duality under imaginary potential}

Consider a uniform chain with imaginary quasiperiodic potential, which has
the form
\begin{equation}
H_{\text{\textrm{Im}}}=\sum_{l=1}^{N}a_{l}^{\dag }a_{l+1}+\mathrm{H.c.}%
+i2\gamma \sum_{l=1}^{N}\mathrm{\cos }(ql)a_{l}^{\dag }a_{l},
\end{equation}%
which is an extension from a standard \textrm{AA} model to a non-Hermitian
version by extending $\gamma $\ to $i\gamma $, and $q=2\pi \beta $, $\beta $
determines the degree of the quasiperiodicity.\ The simplicity of $H_{\text{%
\textrm{Im}}}$ allows for exact theoretical statements in certain cases. For
example, the $1$\textrm{D} disordered Anderson model $H_{\mathrm{And}}$\
with real potential allows for only localized eigenstates at all energies
independent of how weak the disorder may be.\ The standard \textrm{AA} model
with the $1$\textrm{D} incommensurate potential has either all eigenstates
extended or localized depending on the strength of the potential. It is due
to the self-duality of the Hamiltonian, which concludes that an \textrm{AA}
Hamiltonian with irrational $q$ manifests a localization transition at the
self-dual point
\begin{equation}
\left\vert \gamma \right\vert =1,  \label{self}
\end{equation}%
i.e., all eigenstates are extended (localized) when $\left\vert \gamma
\right\vert <1$ ($\left\vert \gamma \right\vert >1$). In the following, it
is readily to show that the same thing happens exactly for a non-Hermitian
\textrm{AA} model.

Taking the Fourier transformation
\begin{equation}
a_{l}=\frac{1}{\sqrt{N}}\sum_{n}(-1)^{l}e^{iqnl}a_{n},
\end{equation}%
we can obtain
\begin{equation}
-iH_{\text{\textrm{Im}}}=\gamma \sum_{n}(a_{n}^{\dag }a_{n+1}+\mathrm{H.c.}%
)+2i\sum_{n}\cos (qn)a_{n}^{\dag }a_{n}.
\end{equation}%
Obviously, $H_{\text{\textrm{Im}}}$ and $-iH_{\text{\textrm{Im}}}$ has the
identical complete set of eigenstates and have self-duality with the
self-dual point $\gamma =1$. It indicates that the imaginary quasiperiodic
potential takes the same role leading to localization transition. Numerical
simulations are given in Fig. \ref{fig3} and Fig. \ref{fig4} to demonstrate
our conclusion about the self-duality under the imaginary potential model.

\section{Interference effect on localization transition}

\label{Interference effect on localization transition}

\begin{figure*}[tbph]
\includegraphics[width=1\textwidth, clip]{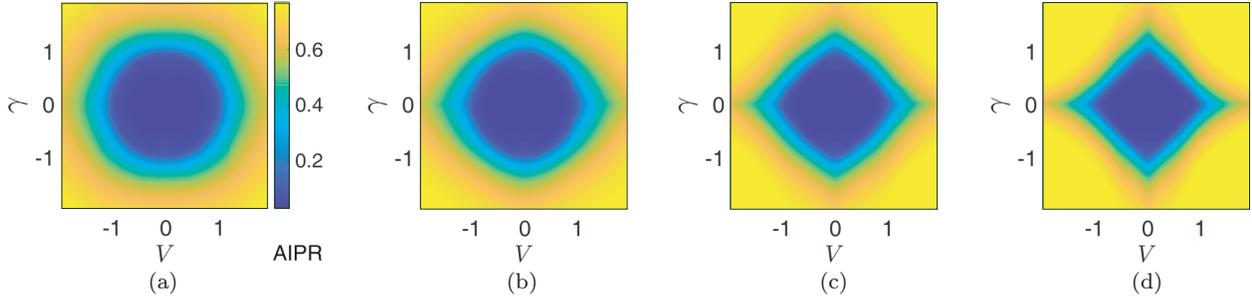}
\caption{(Color online) Color contour maps of the \textrm{AIPR} in the
complex $\protect\gamma -V$ plane for the system of Eq. (\protect\ref{Hcomp}%
)\ with several types of potential configurations. Cause the period of
patterns is $\protect\pi /2$, then the simulations are given for the
potential with four typical phases (a) $\Phi =0$, (b) $\Phi =\protect\pi /16$%
, (c) $\Phi =3\protect\pi /16$, and (d) $\Phi =\protect\pi /2$. We see that
the \textrm{AIPR} along $V$ and $\protect\gamma $\ axes indicate the
localization transition point, which accords with our analysis in Eq. (%
\protect\ref{self}). Beyond the axes, the transition point exhibits evident
interference pattern. The system parameters are $N=500$, $\protect\beta =(%
\protect\sqrt{5}-1)/2$. }
\label{fig3}
\end{figure*}

\begin{figure*}[tbph]
\includegraphics[width=1\textwidth, clip]{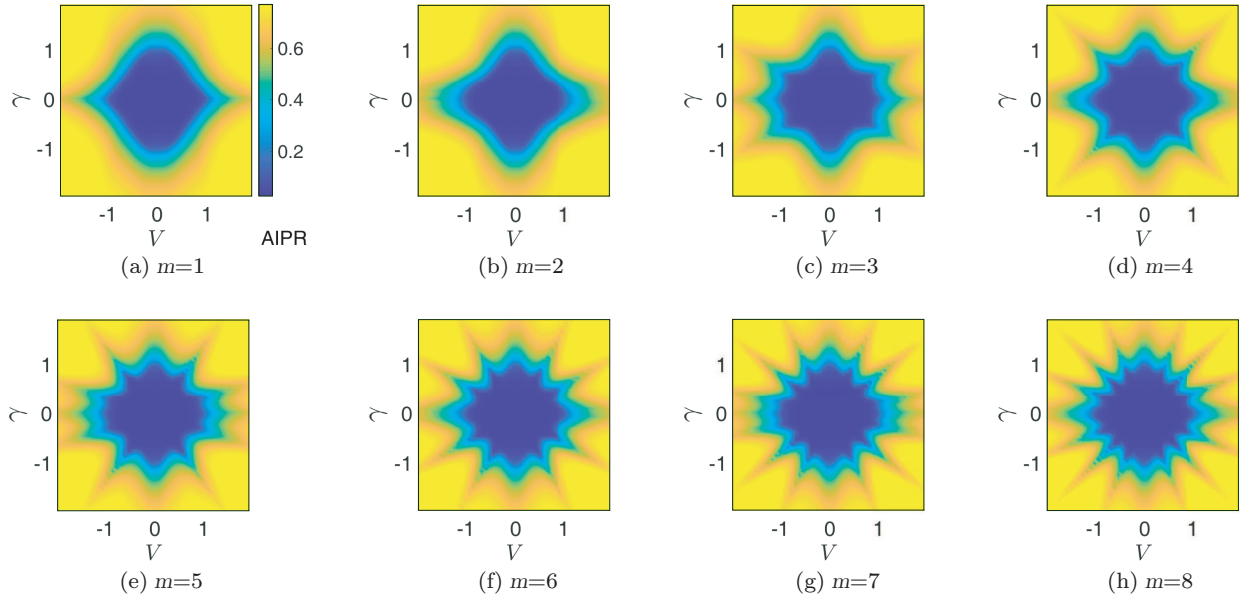}
\caption{(Color online) The same as Fig. \protect\ref{fig3} but for the
system with potential in the expression $\Phi =m\tan ^{-1}(\protect\gamma %
/V) $. Plots of eight typical patterns with integer $m=1,2,3,...,8$,
respectively.}
\label{fig4}
\end{figure*}

We have seen that a real or imaginary quasiperiodic potential has the same
effect on the localization when they show up individually. A natural
question is what happens when the quasiperiodic potential is complex. There
are many variety of complex potential. In this work, we consider the complex
potential with identical frequency
\begin{eqnarray}
H_{\mathrm{Comp}} &=&\sum_{j=1}^{N}(a_{j}^{\dag }a_{j+1}+\mathrm{H.c.}%
)+2V\sum_{j=1}^{N}\mathrm{\cos }(2\pi \beta j)a_{j}^{\dag }a_{j}  \notag \\
&&+i2\gamma \sum_{j=1}^{N}\mathrm{\cos }(2\pi \beta j+\Phi )a_{j}^{\dag
}a_{j},  \label{Hcomp}
\end{eqnarray}%
but different phase difference $\Phi $. In the following, we aim at the
effect of $\Phi $\ on the localization transition point. We consider the
form of $\Phi $\ in two simple ways. (i) We take the $\Phi $ in the range of
$[0,2\pi ]\ $and it is$\ $independent of the values of $V$ and $\gamma $.
(ii) Phase difference $\Phi $ is taken as a function of $V$ and $\gamma $, $%
\Phi =m\tan ^{-1}(\gamma /V)$, where $m$ is an integer. We note that the
point along $V$\ and $\gamma $\ axes of $\gamma -V$\ plane can be exactly
solved as the transition point, i.e., when $|\gamma |<1$\ ($|\gamma |>1$) at
$\gamma $\ axis, all the eigenstates are extended (or localized), the same
as $\gamma $\ axis but for the point $|V|<1$\ ($|V|>1$) at $V$\ axis. The
numerical simulation is performed in the following procedure. For a given
point $(V,\gamma )$, we compute the eigenstates by exact diagonalization.
Then the \textrm{IPR} of all the levels and the average \textrm{IPR} \textbf{%
(}\textrm{AIPR}\textbf{)} for all energy levels can be obtained. Meanwhile,
we confirm that the maximal and minimal \textrm{IPR} has not much deviations
from the average,\textbf{\ }then\textbf{\ }we can use \textrm{AIPR}\ to
describe the characters of extended (or localized) eigenstates.\textbf{\ }In
Figs. \ref{fig3} and \ref{fig4}, we find that the localization transition
boundary is strongly depends on both the magnitude and the phase of the
complex potential. The \textrm{AIPR} profiles exhibit interference-like
patterns in the parameter space. We also notice that patterns in both
figures are asymmetrically under the switch of $V$ and $\gamma $. It is
probably due to the Hermiticity of the hopping term, which breaks the
balance between the real and imaginary potential.

\section{Conclusion}

\label{Conclusion}

In summary, we have investigated an extension of localization to
non-Hermitian system. We have obtained three main results. (i) Based on the
equivalence between a disorder non-Hermitian \textrm{SSH} model and a
Hermitian uniform chain with real random on-site potential, we have shown
exactly that the uncorrelated disorder imaginary on-site potential can
induce Anderson localization. (ii) We presented a non-Hermitian \textrm{AA}
model, by replacing a real quasiperiodic potential by an imaginary one. It
has been shown to have self-duality and thus supports the transition from
extension to localization. (iii) Furthermore, we also studied the
interactive effect of real and imaginary quasiperiodic potential on the
localization transition. Numerical simulations indicates that the phase
difference between real and imaginary quasiperiodic potential determines the
boundary of transition, exhibiting an interference-like pattern. Our
approach opens a new way to investigate the interplay of localization and
gain/loss in non-Hermitian system.

\section*{Appendix}

\label{Appendix}

In this Appendix, we present a detailed derivation for the diagonalization
of a non-Hermitian \textrm{SSH} chain with Hamiltonian
\begin{equation}
H_{\mathrm{SSH}}=(1-\delta )\sum_{j=1}^{N-1}b_{j}^{\dag }a_{j+1}+\mathrm{H.c.%
}+\sum_{j=1}^{N}h_{j}.
\end{equation}%
Here each $h_{j}$\ describes a non-Hermitian $\mathcal{PT}$-symmetric dimer
\begin{equation}
h_{j}=(1+\delta )(a_{j}^{\dag }b_{j}+\mathrm{H.c.})+i\gamma _{j}(a_{j}^{\dag
}a_{j}-b_{j}^{\dag }b_{j}),
\end{equation}%
at position $j$. It can be diagonalized as the form
\begin{equation}
h_{j}=\sqrt{(1+\delta )^{2}-\gamma _{j}^{2}}(\overline{\alpha }_{j}\alpha
_{j}-\overline{\beta }_{j}\beta _{j}),
\end{equation}%
by introducing canonical particle operators in the aid of biorthogonal inner
product. The particle operators
\begin{equation}
\left\{
\begin{array}{cc}
\alpha _{j}=\frac{a_{j}+e^{-i\varphi _{j}}b_{j}}{1+ie^{-i\varphi _{j}}}, &
\beta _{j}=\frac{a_{j}-e^{i\varphi _{j}}b_{j}}{1+ie^{i\varphi _{j}}} \\
\overline{\alpha }_{j}=\frac{a_{j}^{\dag }+e^{-i\varphi _{j}}b_{j}^{\dag }}{%
1-ie^{-i\varphi _{j}}}, & \overline{\beta }_{j}=\frac{a_{j}^{\dag
}-e^{i\varphi _{j}}b_{j}^{\dag }}{1-ie^{i\varphi _{j}}}%
\end{array}%
\right. ,
\end{equation}%
with
\begin{equation}
\tan \varphi _{j}=\frac{\gamma _{j}}{\sqrt{(1+\delta )^{2}-\gamma _{j}^{2}}},
\end{equation}%
or inversely
\begin{equation}
\left\{
\begin{array}{c}
a_{j}=\frac{\left( e^{i\varphi _{j}}+i\right) \alpha _{j}+\left(
e^{-i\varphi _{j}}+i\right) \beta _{j}}{2\cos \varphi _{j}} \\
b_{j}=\frac{\left( 1+ie^{-i\varphi _{j}}\right) \alpha _{j}-\left(
1+ie^{i\varphi _{j}}\right) \beta _{j}}{2\cos \varphi _{j}} \\
a_{j}^{\dag }=\frac{\left( e^{i\varphi _{j}}-i\right) \overline{\alpha }%
_{j}+\left( e^{-i\varphi _{j}}-i\right) \overline{\beta }_{j}}{2\cos \varphi
_{j}} \\
b_{j}^{\dag }=\frac{\left( 1-ie^{-i\varphi _{j}}\right) \overline{\alpha }%
_{j}-\left( 1-ie^{i\varphi _{j}}\right) \overline{\beta }_{j}}{2\cos \varphi
_{j}}%
\end{array}%
\right. ,
\end{equation}%
which satisfy the canonical commutation relations
\begin{eqnarray}
\left[ \alpha _{j},\overline{\alpha }_{j^{\prime }}\right] _{\pm } &=&\left[
\beta _{j},\overline{\beta }_{j^{\prime }}\right] _{\pm }=\delta
_{jj^{\prime }}, \\
\left[ \alpha _{j},\alpha _{j^{\prime }}\right] _{\pm } &=&\left[ \beta
_{j},\beta _{j^{\prime }}\right] _{\pm }=\left[ \overline{\alpha }_{j},%
\overline{\alpha }_{j^{\prime }}\right] _{\pm }=\left[ \overline{\beta }_{j},%
\overline{\beta }_{j^{\prime }}\right] _{\pm }=0,  \notag
\end{eqnarray}%
for the interchain particles,
\begin{equation}
\left[ \alpha _{j},\overline{\beta }_{j^{\prime }}\right] _{\pm }=\left[
\overline{\alpha }_{j},\overline{\beta }_{j^{\prime }}\right] _{\pm }=\left[
\alpha _{j},\beta _{j^{\prime }}\right] _{\pm }=\left[ \overline{\alpha }%
_{j},\beta _{j^{\prime }}\right] _{\pm }=0,
\end{equation}%
for the intrachain particles. Straightforward derivations show that
\begin{eqnarray}
&&(1+\delta )(a_{j}^{\dag }b_{j}+b_{j}^{\dag }a_{j})+i\gamma
_{j}(a_{j}^{\dag }a_{j}-b_{j}^{\dag }b_{j})  \notag \\
&=&\sqrt{(1+\delta )^{2}-\gamma _{j}^{2}}(\overline{\alpha }_{j}\alpha _{j}-%
\overline{\beta }_{j}\beta _{j}),
\end{eqnarray}%
and%
\begin{align}
& b_{j}^{\dag }a_{j+1}+a_{j+1}^{\dag }b_{j}  \notag \\
& =[\left( e^{i\varphi _{j+1}}+i\right) \left( 1-ie^{-i\varphi _{j}}\right)
\overline{\alpha }_{j}\alpha _{j+1}  \notag \\
& +\left( e^{-i\varphi _{j+1}}+i\right) \left( 1-ie^{-i\varphi _{j}}\right)
\overline{\alpha }_{j}\beta _{j+1}  \notag \\
& -\left( e^{i\varphi _{j+1}}+i\right) \left( 1-ie^{i\varphi _{j}}\right)
\overline{\beta }_{j}\alpha _{j+1}  \notag \\
& -\left( e^{-i\varphi _{j+1}}+i\right) \left( 1-ie^{i\varphi _{j}}\right)
\overline{\beta }_{j}\beta _{j+1}]\times  \notag \\
& (4\cos \varphi _{j}\cos \varphi _{j+1})^{-1}+\mathrm{H.c.}.
\end{align}%
Under the condition
\begin{equation}
1+\delta \gg 1-\delta ,\left\vert \gamma _{j}\right\vert \ll 1,
\end{equation}%
we have
\begin{eqnarray}
\cos \varphi _{j} &=&\frac{\sqrt{(1+\delta )^{2}-\gamma _{j}^{2}}}{1+\delta }%
\approx 1-\frac{\gamma _{j}^{2}}{2(1+\delta )^{2}}\approx 1, \\
e^{i\varphi _{j}} &=&\frac{\sqrt{(1+\delta )^{2}-\gamma _{j}^{2}}+i\gamma
_{j}}{1+\delta }\approx 1+\frac{i\gamma _{j}}{1+\delta }\approx 1,
\end{eqnarray}%
and then
\begin{eqnarray}
&&b_{j}^{\dag }a_{j+1}+a_{j+1}^{\dag }b_{j}  \notag \\
&\approx &\frac{1}{2}(\overline{\alpha }_{j}\alpha _{j+1}+\overline{\alpha }%
_{j}\beta _{j+1}-2\overline{\beta }_{j}\alpha _{j+1}-\overline{\beta }%
_{j}\beta _{j+1})  \notag \\
&&+\frac{1}{2}(\alpha _{j+1}^{\dag }\overline{\alpha }_{j}^{\dag }-\alpha
_{j+1}^{\dag }\overline{\beta }_{j}^{\dag }+\beta _{j+1}^{\dag }\overline{%
\alpha }_{j}^{\dag }-\beta _{j+1}^{\dag }\overline{\beta }_{j}^{\dag }) \\
&=&\frac{1}{2}(\overline{\alpha }_{j}\alpha _{j+1}+\overline{\alpha }%
_{j}\beta _{j+1}-\overline{\beta }_{j}\alpha _{j+1}-\overline{\beta }%
_{j}\beta _{j+1})+\mathrm{H.c.}.  \notag
\end{eqnarray}%
Neglecting the transition terms between sites with opposite potential, we
have
\begin{equation}
b_{j}^{\dag }a_{j+1}+a_{j+1}^{\dag }b_{j}\approx \frac{1}{2}\overline{\alpha
}_{j}\alpha _{j+1}-\frac{1}{2}\overline{\beta }_{j}\beta _{j+1}+\mathrm{H.c.}%
.
\end{equation}%
Then we have the approximate expression
\begin{eqnarray}
H_{\mathrm{SSH}} &\approx &\frac{1}{2}(1-\delta )\sum_{j=1}^{N-1}\left(
\overline{\alpha }_{j}\alpha _{j+1}-\overline{\beta }_{j}\beta _{j+1}\right)
\notag \\
&&+\mathrm{H.c.}+\sum_{j=1}^{N}V_{j}(\overline{\alpha }_{j}\alpha _{j}-%
\overline{\beta }_{j}\beta _{j}),
\end{eqnarray}%
where real potential $V_{j}=\sqrt{(1+\delta )^{2}-\gamma _{j}^{2}}$. On the
other hand, we have
\begin{eqnarray}
\left[ \alpha _{j},\alpha _{j^{\prime }}^{\dag }\right] _{\pm } &=&\frac{%
\delta _{jj^{\prime }}}{1+\sin \varphi _{j}},\left[ \beta _{j},\beta
_{j^{\prime }}^{\dag }\right] _{\pm }=\frac{\delta _{jj^{\prime }}}{1-\sin
\varphi _{j}}, \\
\left[ \alpha _{j},\alpha _{j^{\prime }}\right] _{\pm } &=&\left[ \beta
_{j},\beta _{j^{\prime }}\right] _{\pm }=\left[ \alpha _{j}^{\dag },\alpha
_{j^{\prime }}^{\dag }\right] _{\pm }=\left[ \beta _{j}^{\dag },\beta
_{j^{\prime }}^{\dag }\right] _{\pm }=0,  \notag
\end{eqnarray}%
for the interchain particles,
\begin{eqnarray}
\left[ \alpha _{j},\beta _{j^{\prime }}^{\dag }\right] _{\pm } &=&i\tan
\varphi _{j}\delta _{jj^{\prime }},\left[ \alpha _{j}^{\dag },\beta
_{j^{\prime }}\right] _{\pm }=-i\tan \varphi _{j}\delta _{jj^{\prime }},
\notag \\
\left[ \alpha _{j},\beta _{j^{\prime }}\right] _{\pm } &=&\left[ \alpha
_{j}^{\dag },\beta _{j^{\prime }}^{\dag }\right] _{\pm }=0,
\end{eqnarray}%
for the intrachain particles. It indicates that if a state satisfy the\
relation
\begin{equation}
\left\vert \psi \right\rangle =\sum_{j}(A_{j}\alpha _{j}^{\dag }\left\vert
0\right\rangle +B_{j}\beta _{j}^{\dag }\left\vert 0\right\rangle ),
\end{equation}%
\ with $A_{j}B_{j}=0$ for all $j$, i.e., it has only single-chain component,
$\left\{ \alpha _{j}^{\dag }\left\vert 0\right\rangle \right\} $ or $\left\{
\beta _{j}^{\dag }\left\vert 0\right\rangle \right\} $, the dynamics of $%
\left\vert \psi \right\rangle $\ is the same as that in a Hermitian chain.

\acknowledgments We acknowledge the support of NSFC (Grants No. 11874225).

\end{document}